\pdfoutput=1
\documentclass[aps,prl,reprint,amsmath,amssymb,nofootinbib,superscriptaddress]{revtex4-1}
\usepackage{custom}
\begin{document}

\title{Universality, Lee-Yang singularities and series expansions}

\author{G\"{o}k\c{c}e Ba\c{s}ar}
\email{gbasar@unc.edu}
\affiliation{Department of Physics and Astronomy, University of North Carolina, Chapel Hill, North Carolina 27599, USA}

\date{\today}

\begin{abstract}   We introduce a new way of reconstructing the equation of state of a thermodynamic system near a second order critical point from a finite set of Taylor coefficients computed away from the critical point. We focus on the Ising universality class (${\mathbb Z}_2$ symmetry) and show that in the crossover region of the phase diagram it is possible to efficiently extract the location of the nearest thermodynamic singularity, the Lee-Yang edge singularity, from which one can (i) determine the location of the critical point, (ii) constrain the non-universal parameters that maps the equation of state to that of the Ising model in the scaling regime, and (iii) numerically evaluate the equation of state in the vicinity of the critical point. This is done by using a combination of Pad\'e resummation and conformal maps. We explicitly demonstrate these ideas in the celebrated Gross-Neveu model. 
\end{abstract}

\maketitle

\section{Introduction}
\label{sec:intro}
In the vicinity of a critical point, the correlation length of a thermodynamic system grows, the underlying microscopic properties become irrelevant, and consequently many different substances exhibit the same behavior. This powerful notion of universality allows one to classify second order phase transitions based on the underlying symmetries of the system without detailed knowledge of the microscopic dynamics. Some famous examples are the liquid-gas and ferromagnetic transitions, both of which possess critical points that are in the (static) universality class of the 3d Ising model. Universality identifies the singular behavior of the equation of state, say pressure as a function of temperature and chemical potential $p(T,\mu)$, with the Ising equation of state as a function of the Ising variables, i.e. the reduced temperature, $r$, and the magnetic field, $h$, near the critical point. However neither the location of the critical point, nor the relation between $(T,\mu)$ and $(r,h)$ are universal, and for a quantitative description of the critical phenomena they have to be computed from the microscopic dynamics of the theory of interest. This is in many cases a mathematically intractable task, and often the solution can only be obtained as a series expansion at a point away from the critical point with a finite number of terms \cite{Fisher:1974series}. 

In this work we consider the following general problem: given a truncated local series expansion of the equation of state in some parameter such as $\mu$, obtained away from the critical point, what can we say about the critical behavior of the system? More precisely, how can we determine whether a critical point exists, and if it does how can we reconstruct the singular behavior of the equation of state near the critical point from the truncated expansion obtained away from it? We show that it is possible to obtain a surprisingly large amount of information about the critical behavior of the system from the series coefficients, even if we have access to a modest number of them.  

A major motivation for this work is the search for the conjectured critical point in the phase diagram of Quantum Chromo Dynamics (QCD) \cite{Stephanov:2004wx} which is one of the major outstanding problems in nuclear physics both theoretically and experimentally \cite{Bzdak:2019pkr}. The theoretical approaches are severely limited by the sign problem that prevents first-principle lattice computations at nonzero baryon chemical potential, $\mu_B$ and one of the methods to deal with this obstacle is to expand the equation of state around $\mu_B=0$ and compute the Taylor coefficients on the lattice without a sign problem (see \cite{Ratti:2018ksb} for a recent review). Another motivation is to understand the properties of strongly interacting fermions, such as unitary Fermi gases. where our theoretical knowledge of the equation of state is typically limited to the first few terms in the virial expansion \cite{Liu_2013}.

\section{Lee-Yang edge singularities }
\label{sec:ly}
Before detailing our method we briefly summarize the Lee-Yang edge singularities that play a crucial role in our analysis. 
In their seminal work phase transitions \cite{Yang:1952be,Lee:1952ig} Lee and Yang showed that the thermodynamic properties of a system is encoded in the distribution of the zeroes of the partition function $Z(\zeta)$ as a function of fugacity, $\zeta=e^{\mu/T}$. For a finite system the partition function is a positive polynomial for $\zeta \geq 0$. However in general it has zeroes for complex values of $\mu$ and $T$. In the thermodynamic limit the zeroes coalesce into branch cuts emanating from the so called Lee-Yang (LY) edge singularities. When the LY singularities pinch the real axis, the system exhibits a second order phase transition. Likewise the branch cut associated with a LY singularity crosses the real line when there is a first order phase transition.

The LY singularities are critical points themselves and have their own universality class. For example, for the Ising model, and $O(N)$ models in general, they are characterized by the $\phi^3$ theory with a pure imaginary coupling \cite{Fisher:1978pf}. For these models the analytical structure of the LY singularities has been studied \cite{An:2017brc,An:2017rfa}. In the complex plane of the scaling variable, $x=h r^{-\bd}$\footnote{$\beta$ and $\delta$ are the standard Ising critical exponents \cite{ZinnJustin:2002ru}.}, the singularities are located in the pure imaginary axis, $x=\pm i x_{LY}$ where $x_{LY}\in {\mathbb R}$ has recently been calculated via the functional renormalization group method \cite{Connelly:2020gwa}. The magnetization around the LY singularity behaves as $m-m_c\sim (x\pm i x_{LY})^\sigma$ with the critical exponent $\sigma\approx0.074-0.085$ for $d=3$ \cite{An:2016lni}. The LY singularities in the context of QCD critical point has been discussed in, for example, \cite{Halasz_1997,Ejiri:2005ts,Stephanov:2006dn,WAKAYAMA2019227,Mukherjee:2019eou,Connelly:2020pno,Schmidt:2021pey}.

Consider the equation of state of a theory near a critical point, $(T_c,\mu_c)$, in the Ising universality class. From universality we can relate it to the Ising model via a linear map \cite{Parotto:2018pwx,Pradeep:2019ccv} 
\bea
\label{eq:mapping}
\begin{pmatrix}
r\\ h
\end{pmatrix}:={\mathbb M}\begin{pmatrix}
T-T_c \\ \mu-\mu_c
\end{pmatrix}=
\begin{pmatrix}
r_T && h_T  
\\
r_\mu && h_\mu 
\end{pmatrix}
\begin{pmatrix}
T-T_c \\ \mu-\mu_c
\end{pmatrix}.\quad
\ea
This relation then leads to the following expression for the trajectory of the LY singularities \cite{Stephanov:2006dn}: 
\bea
\label{eq:ly-traj}
\mu_{LY}(T)\approx\mu_c -c_1(T-T_c) \pm i x_{LY} c_2 (T-T_c)^\bd, 
\nn
\text{where }c_1:=\frac{\hT}{\hmu} \quad c_2:=\frac{\rmu^\bd}{\hmu } \prt{\frac{\rT}{\rmu}-\frac{\hT}{\hmu} }^{\bd}.\quad
\ea
Notice that $c_1$ is the slope of the crossover line, whereas $c_2$ depends on the relative angle between the $h$ and $r$ axes \cite{Parotto:2018pwx, Pradeep:2019ccv}. Therefore the trajectory in Eq. \eqref{eq:ly-traj} depends on not only the location of the critical point, but also on the non-universal mapping parameters. We now explain how to optimally construct Eq. \eqref{eq:ly-traj} from a series expansion. 

\section{The method}
\label{sec:method}   
  Consider a thermodynamic function, $f(T,\mu)$, (pressure, density, susceptibility etc.) given a series expansion with finite number of terms: $f(T,\mu)\sim\sum_{n=0}^N f_n(T)\mu^{2n}$. Our goal is to extract as much information from it as possible, especially about its singular behavior near a critical point if there is one. Obviously, in this form what we have is a polynomial which does not exhibit any singular behavior. In principle from a ratio test it is possible determine the radius of convergence which would indicate the location of the closest singularity to origin. However when the nearest singularities are a complex conjugate pair of LY singularities (which is the case for a smooth crossover)  the ratios of series coefficients do not converge monotonically but rather have an oscillating envelope as a result of Darboux's theorem\footnote{For some estimators for the radius of of convergence for QCD see, i.e. \cite{PhysRevD.99.114510,GIORDANO2021121986}.}. This makes numerically extracting the singularity from the ratios challenging. 
Alternatively, the singular behavior of $f(T,\mu)$ can be approximately constructed by a Pad\'e resummation, ${\rm P_{N/2}}[f](\mu^2):=p(\mu^2)/q(\mu^2)$ where $p$ and $q$ are polynomials of order $N/2$ whose the coefficients are determined by expanding ${\rm P_{N/2}}[f](\mu^2)$ and identifying the coefficients with the Taylor coefficients.  The singularities of $f$, typically branch points, are approximated by the poles and zeroes of the Pad\'e approximant. However as we will demonstrate later, Pad\'e resummation   
has a known shortcoming; it creates spurious singularities which limits its range of applicability. In addition, it is also not the most optimal approximation scheme and can be dramatically improved by pairing with a conformal map \cite{Costin:2020pcj}. The key idea is to apply an appropriately chosen conformal map, $\mu^2:=\phi(z)$, and do the Pad\'e reummation in $z$: ${\rm CP_{N/2}}[f](z):=\tilde p(z)/\tilde q(z)$ where $\tilde p$ and $\tilde q$ are order $N/2$ polynomials whose coeffcients are determined by the Taylor coefficients of $f(\phi(z))$.  The singularities of ${\rm CP_{N/2}}[f](z)$  are mapped to the complex $\mu^2$ plane as via $\phi(z)$. For the remainder of the letter we shall refer to this method of extracting singularities simply as ``conformal Pad\'e". Unlike Pad\'e, Conformal Pad\'e does not suffer from the spurious pole problems. It also provides an optimal approximation to the original function for a wide range of functions as proven in \cite{Costin:2020pcj}. Our strategy is to first construct the trajectory \eqref{eq:ly-traj} by extracting the $\mu_{LY}(T)$ via conformal Pad\'e for a sequence of temperatures and then to obtain the location of the critical point as well as the coefficients $c_1$, $c_2$ from it. This is possible since the critical exponents are fixed by universality and, as mentioned, $x_{LY}$ known \cite{Connelly:2020gwa}.
  
Conformal Pad\'e methods are typically used to reconstruct Borel plane singularities in resumming asymptotic series, such as the $\epsilon$ expansion \cite{Guida:1998bx} or perturbation series for relativistic \cite{Serone:2019szm} and non-relativistic \cite{Rossi:2018} systems. Here we take a different approach and apply it to a convergent series to directly extract its singular behavior near the critical point. Our input, the series expansion of the equation of state, does not have to come from perturbation theory or the $\epsilon$ expansion. Our approach is similar to that of \cite{Fisher:1974series} but with a key differences: we focus on the complex LY singularities and conformal maps play a crucial role in reconstructing the equation of state.

\section{The Gross-Neveu Model}
\label{sec:GN}
To concretely demonstrate these ideas we focus on the celebrated Gross-Neveu (GN) model \cite{Gross:1974jv} which is a four-fermion theory with the action
\bea
\label{eq:GN-lag}
S=\int d^2x \prt{ i\bpsi(\slashed\del -m_q )\bpsi+ \frac{g^2}{2} (\bpsi \psi)^2 },
\ea
where $\psi$ is a Dirac fermion with $N_f$ flavors. It exhibits some of the key features of QCD, such as asymptotic freedom, chiral symmetry breaking and dimensional transmutation. Notably it was also shown that the GN model gives a reasonably good description of the first order phase transition in doped polyacetlyne  \cite{Chodos:1993mf}. The theory has a discrete (${\mathbb Z}_2$) chiral symmetry, $\psi\rightarrow \gamma_5\psi$, for $m_q=0$. We will work in the large $N_f$ limit $N_f\rightarrow \infty$ with $g^2N_f$=fixed, where the fluctuations are suppressed and the mean field solution is exact. 

The exact phase diagram of the GN model is known \cite{Schnetz:2004vr,Schnetz:2005ih} and has a rich structure such as spatially inhomogeneous kink/anti-kink crystals at high densities \cite{Schnetz:2005ih} and an exactly soluble, all-orders Ginzburg Landau expansion \cite{Basar:2009fg}. However, in order to keep the discussion simple, we shall assume that the translational symmetry remains unbroken  in this work. Furthermore we focus on the crossover region of the phase diagram which is not affected by the existence of inhomogeneous phases. 

\begin{figure}
\includegraphics[scale=0.37]{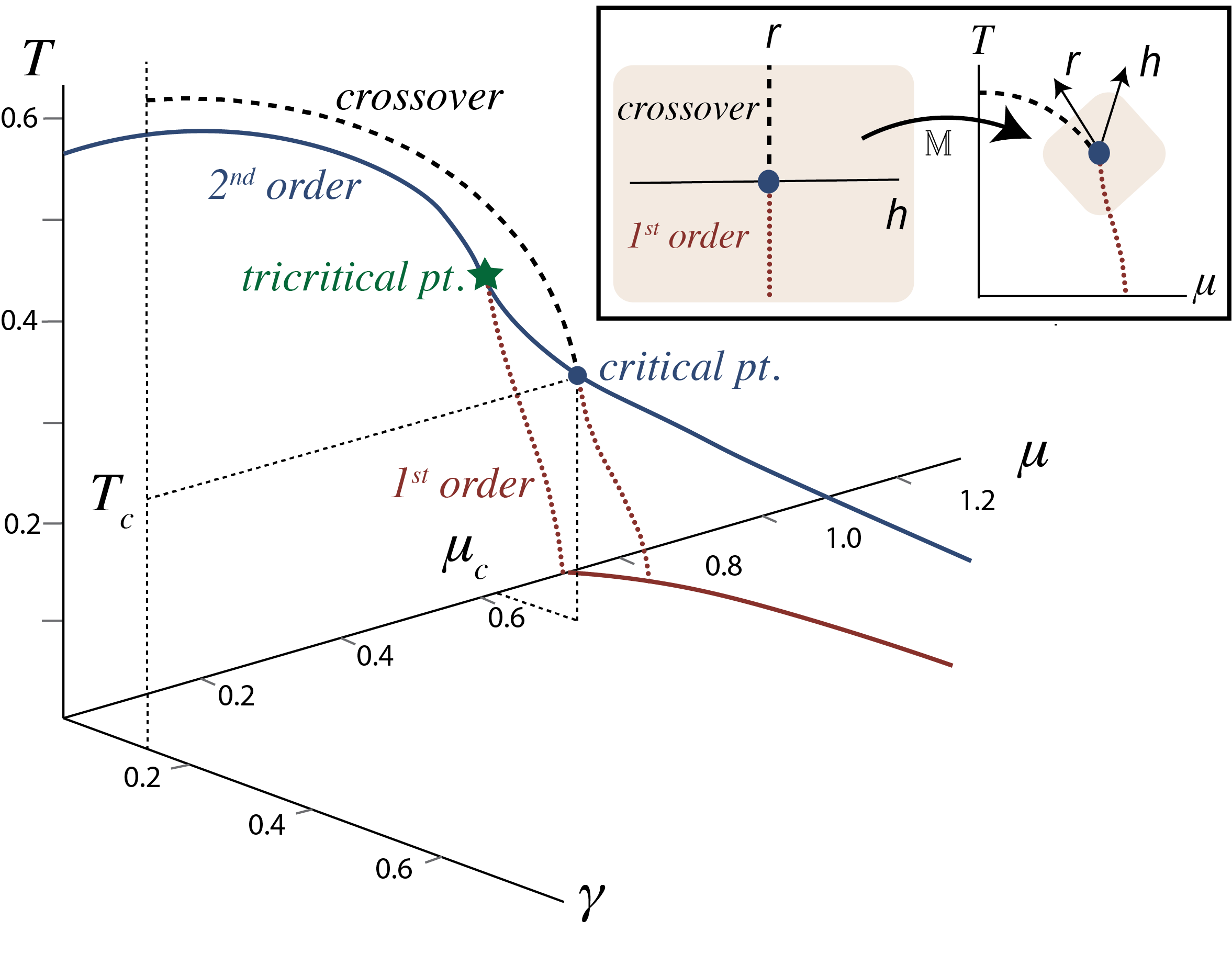}
\caption{The phase diagram of the Gross-Neveu model assuming unbroken translational symmetry. 
Inset: The mapping between the Ising model parameters, $r,h$, and $T,\mu$ near the critical point given in Eq. \eqref{eq:mapping}.
}
\label{fig:GN-phase}
\end{figure}

Thermodynamics of the model follows from minimizing the grand potential 
\bea
\label{eq:GN-free}
\Omega(\phi)&=&\frac{\phi^2}{ 2\pi}\left(\log \phi-\frac12+\gamma\right)-\frac{\gamma}{ \pi}\phi \\
&&-T \int \frac{d k}{ 2\pi} \prod_{\eta=\pm1}\log \sqprt{\prt{1+e^{-({\sqrt{k^2+\phi^2}+\eta\mu})/T}} }
\nonumber
\ea
with respect to $\phi$ which determines the fermion mass, $M$ as a function of $T,\mu$ and $\gamma$. The parameter $\gamma\propto m_q$ is a measure of explicit chiral symmetry breaking and it vanishes in the chiral limit \cite{Schnetz:2004vr}. In our analysis we will work with a fixed, nonzero value of $\gamma$.
The equation of state is obtained by identifying the pressure as $p(T,\mu)=-\Omega[M(T,\mu)]:=-{\rm min}_\phi\Omega[\phi]$. The homogeneous phase diagram of the model is shown in Fig. \ref{fig:GN-phase}. In the chiral limit ($\gamma=0$) the ordered phase, $M\neq0$, where chiral symmetry is broken, and the disordered phase, $M=0$, where the chiral symmetry is restored are separated by a second (first) order transition for $T<T_{tc}$  ($T>T_{tc}$)
shown in solid blue (dotted red) curves. These curves merge at a tricritical point $(T_{tc},\mu_{tc})\approx(0.318,0.608)$\footnote{All dimensionful quantities in this letter, such as $T$ and $\mu$, are expressed in units of the vacuum fermion mass.}. When the chiral symmetry is explicitly broken, for a fixed $\gamma\neq0$, the transition is a smooth crossover for hight $T$ that ends at a critical point, $(T_c,\mu_c)$, and turns into a first order transition.
Since the theory has a ${\mathbb Z}_2$ chiral symmetry the second order transition is in the same static universality class as the mean field Ising model. The trajectory of the LY singularities, $\mu_{LY}(T)$, is determined by the condition 
\bea
\label{eq:LY-cond}
\partial_\phi \Omega (\phi)=\partial_\phi^2  \Omega (\phi)=0\,.
\ea
In the crossover region, $T>T_c$, this condition leads to a pair of complex solutions $\mu_{LY}(T)={\rm Re} \mu_{LY}(T) \pm i {\rm Im} \mu_{LY}(T)$ which coalesce and pinch the real axis at the ordinary critical point as expected: $\mu_{LY}(T_c)=\mu_c$.  In the vicinity of the critical point $\mu_{LY}(T)$ takes scaling form given in Eq. \eqref{eq:ly-traj} with the mean field exponents, $\beta=1/2,\delta=3$:  
\bea
\label{eq:ly-traj-gn}
\mu_{LY}(T)\sim \mu_c- c_1 (T-T_c)+i c_2 x_{LY}(T-T_c)^{3/2}\quad
\ea
 where in the mean field limit $x_{LY}=2/3\sqrt{3}$. In the next section we compute $T_c$, $\mu_c$, $c_1$, $c_2$ directly from a truncated series expansion of the equation of state and compare these results with the exact solution obtained by numerically solving Eq. \eqref{eq:LY-cond}.   

\section{Results}
\begin{figure}
\includegraphics[scale=0.56]{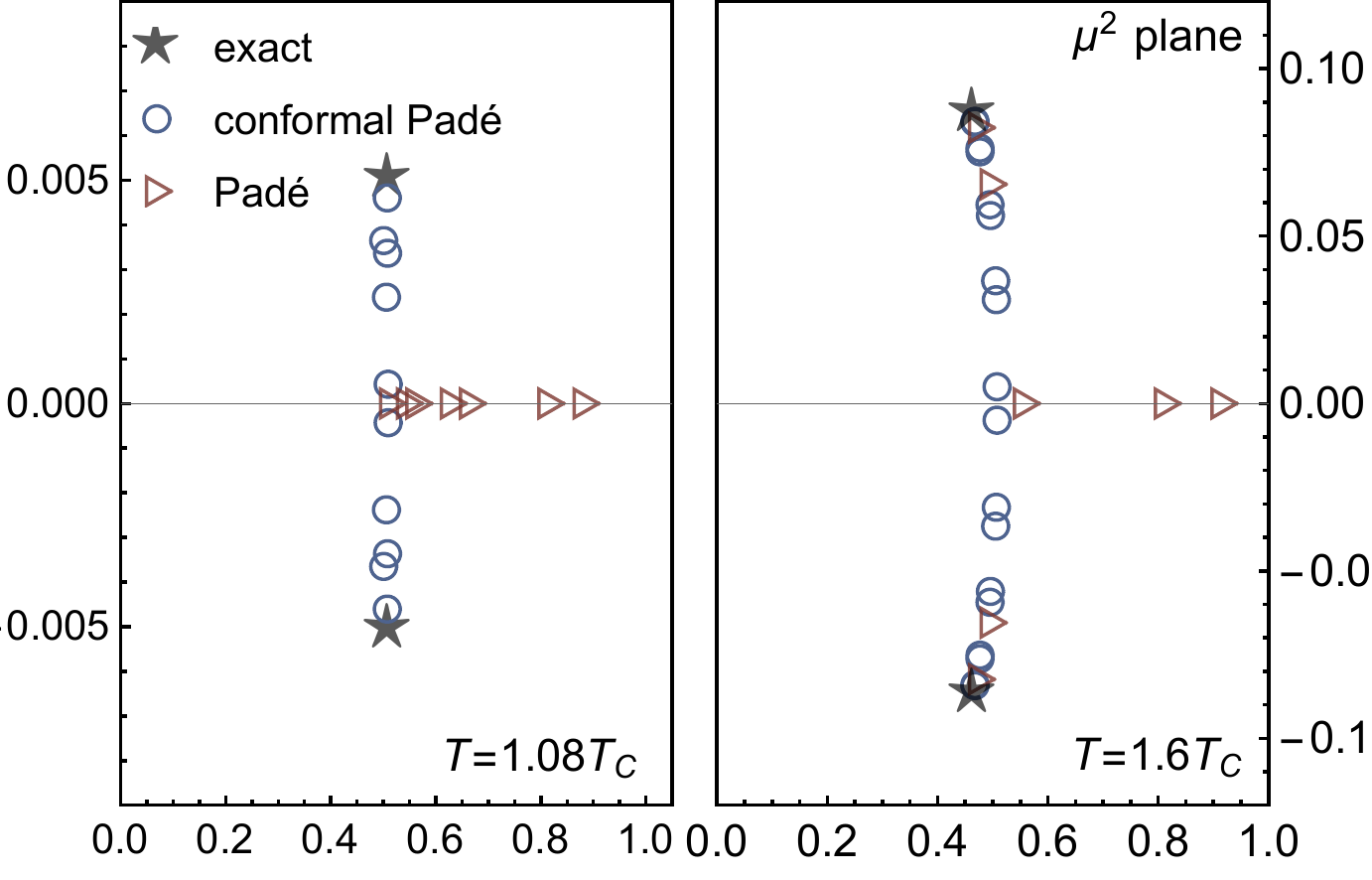}
\caption{The poles and zeroes of Pad\'e and conformal Pad\'e approximants for two different temperatures compared with the exact locations of $\mu_{LY}^2$. }
\label{fig:poles}
\end{figure}

We computed the equation of state perturbatively in $\mu^2$ by first solving $\partial_\phi \Omega(\phi)=0$ order-by-order for a range of temperatures with $\gamma=0.1$. By plugging this solution into Eq. \eqref{eq:GN-free} and expanding in $\mu^2$ we obtained the Taylor series expansion for the pressure $p(T,\mu)\approx\sum_{n=0}^N p_{2n}(T)\mu^{2n}$. 
To illustrate the numerical evaluation of the equation of state we focus on the susceptibility,
\bea
\chi(T,\mu)=\frac{\del^2 p}{ \del \mu^2}\approx \sum_{n=0}^{N-1}  (2n+2)(2n) p_{2n+2}(T)\mu^{2n},\quad
\ea
 because its singular part in the vicinity of the critical point it grows as $\chi(\mu)\sim{\rm Re} (\mu^2-\mu_{LY}^2)^{\sigma-1}$ where $\sigma=1/2$ in the mean field limit. Of course, in many cases it is very difficult, to generate such large number of terms. Therefore we also show results obtained by 11 terms for comparison. We computed the singularities both from Pad\'e and conformal Pad\'e which are shown in  Fig. \ref{fig:poles} for two different temperatures very close to and away from the critical point.  We used a simple conformal map, 
\bea
\label{eq:phi1}
\phi_1(z)=\frac{4\mu_{LY}^2 z}{ (1+z)^2},
\ea
defined over one-cut complex plane with a singularity located at $\mu_{LY}^2$ to resolve $f$ near the singularity $\mu_{LY}^2$. Since the other singularity is the complex conjugate pair, $\mu_{LY}^{*2}$, including its contribution is trivial. Of course, a priori, we don't know what $\mu_{LY}$. Therefore we first obtained a crude estimate for it from regular Pad\'e and we used it in $\phi_1(z)$,  and refined this estimate via conformal Pad\'e.

\begin{figure}
\includegraphics[scale=0.52]{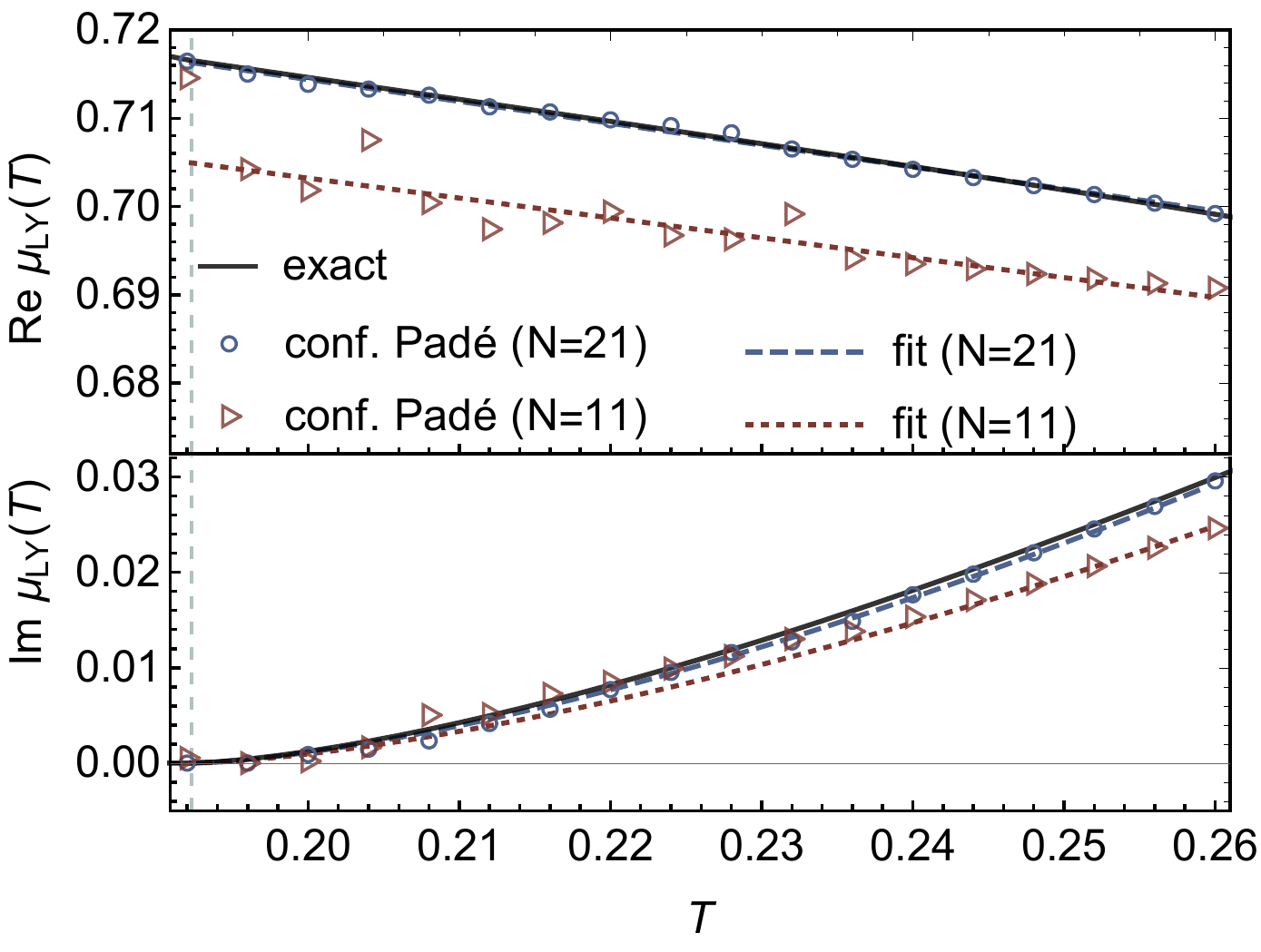}
\caption{The Lee-Yang singularity trajectory, $\mu_{LY}(T)$, reconstructed from conformal Pad\'e with 20 and 10 terms. The vertical line denotes $T_c$.}
\label{fig:ly-traj}
\end{figure}

  In order to reconstruct the trajectory, $\mu_{LY}(T)$, we repeated this procedure for different temperatures. In order to smooth out the $T$ dependence of $\mu_{LY}$ we used fits whose forms are determined from Eq. \eqref{eq:ly-traj-gn}; namely a linear fit for ${\rm Re}\mu_{LY}(T)$ and a $y= a x^{3/2}$ fit for ${\rm Im}\mu_{LY}(T)$. The results are shown in Fig. \ref{fig:ly-traj}. From these fits we obtained the values of $T_c,\mu_c, c_1$ and $c_2$ shown in Table \ref{table}.  
\begin{table}
\begin{tabular}{  | c  | c | c | c |  c | } 
 \hline
 \, & $T_c$ & $\mu_c$ & $c_1$ & $c_2$ 
\\
\hline
 exact & 0.192 & 0.717 &  0.249 & 4.684
 \\ 
conf. Pad\'e (N=21) &  0.195 & 0.716 &0.248  & 4.323
 \\ 
conf. Pad\'e (N=11)& 0.185 & 0.707 & 0.225 & 3.666
\\
 \hline
\end{tabular}
 \caption{The location of the critical point and the Ising model mapping parameters given in Eq. \eqref{eq:ly-traj-gn} extracted from conformal Pad\'e.}
 \label{table}
\end{table}

\begin{figure}
\includegraphics[scale=0.55]{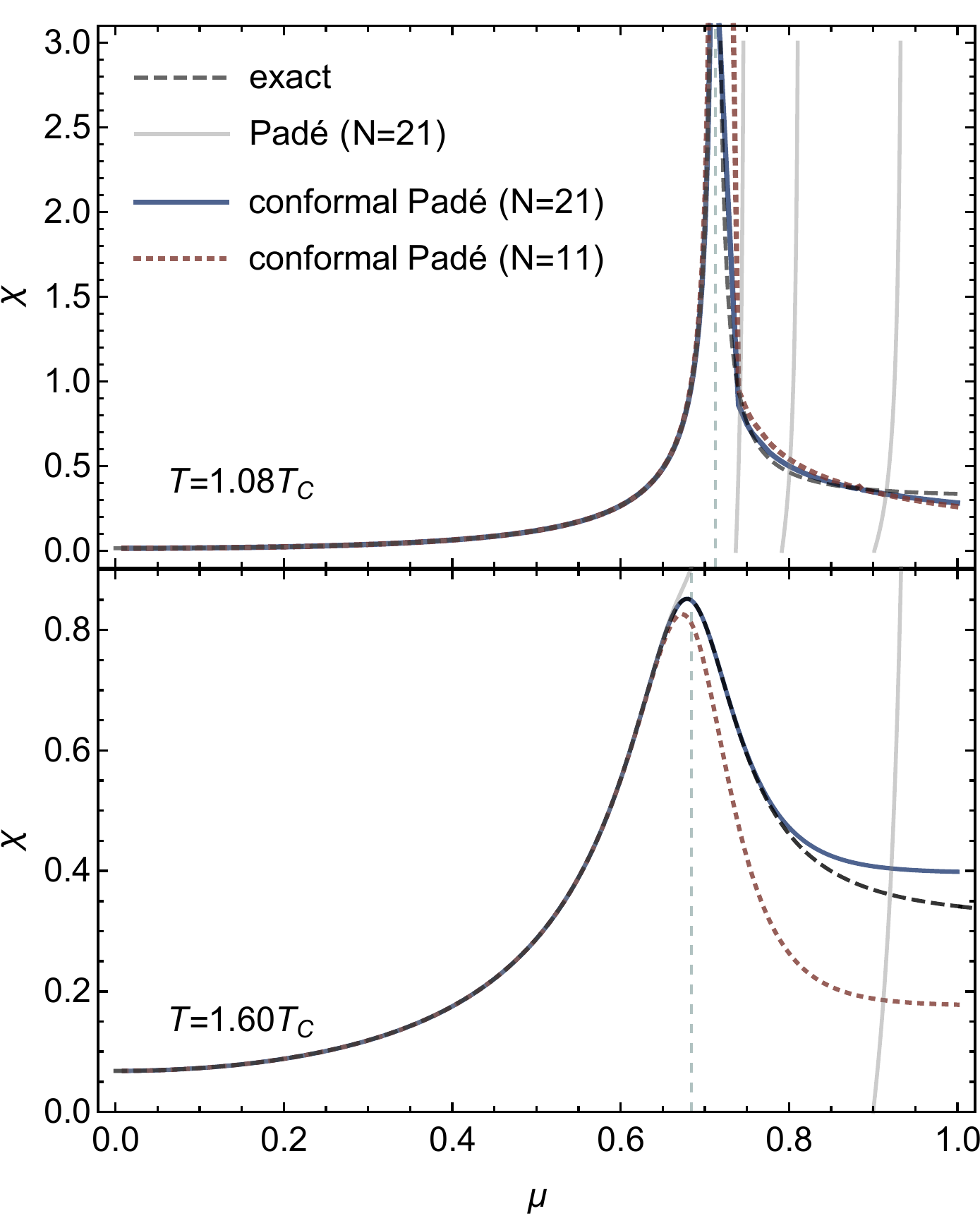}
\caption{The susceptibility as a function of $\mu$ for two different temperatures. Vertical lines denote ${\rm Re}\mu_{LY}(T)$.}
\label{fig:sus}
\end{figure}

Finally we computed the susceptibility as a function $\mu$ via Pad\'e and conformal Pad\'e. In order to capture the global behavior of the equation of state we used a different conformal map,
\bea
\label{eq:phi2}
\phi_2(z)=4|\mu_{LY}|^2\left[\frac {\theta}{(1-z)^2}\right]^\theta\left[\frac {1-\theta}{(1+z)^2}\right]^{1-\theta},\quad 
\ea
defined on a two-cut complex plane with two branch points located at $|\mu_{LY}|^2 e^{\pm  i \pi \theta}$ \cite{Rossi:2018,Serone:2019szm,Costin:2020pcj}. The results for two representative temperatures near and away from $T_c$ are shown in Fig \ref{fig:sus}. 

\section{Discussion}
\label{sec:discussion}
We now discuss these results. Firstly, notice that near the critical point, $T=1.08T_c$, Pad\'e cannot resolve ${\rm Im}\mu_{LY}$ and creates a sequence of poles/zeroes along the real axis. Even when it does, away from the critical point, $T=1.60T_c$, there are still spurious poles along the real axis which makes Pad\'e useless for $\mu \gtrsim {\rm Re}\mu_{LY}$ as seen in Fig. \ref{fig:sus} (gray curves). As seen from the same figures, conformal Pad\'e does not suffer from such a problem and even for 11 terms, it does capture the qualitative behavior of the peak.  For 21 terms the agreement with the exact result up to $\mu \approx 0.8$ is quite remarkable. As mentioned above, in order to refine the location of a given singularity, $\mu_{LY}$, we found the one-cut conformal map given in Eq.\eqref{eq:phi1}  to be very accurate and easy to execute numerically. However globally it generates singularities between $\mu_{LY}^2$  and $\mu_{LY}^{*2}$ as seen in Fig. \ref{fig:poles}. That is why we used a two-cut map given in Eq. \eqref{eq:phi2} to evaluate $\chi(T,\mu)$ shown in Fig. \ref{fig:sus}.

From the singularities extracted from conformal Pad\'e we constructed the LY singularity trajectory Eq. \eqref{eq:ly-traj-gn}. The critical temperature, $T_c$, is obtained from the point ${\rm Im} \mu_{LY}$ vanishes (the vertical line in Fig. \ref{fig:ly-traj}). Therefore it is crucial to be able to accurately compute ${\rm Im} \mu_{LY}$ where conformal Pad\'e  has a significant advantage over other methods, even with an input of 11 Taylor coefficients. The remaining parameters $\mu_c$, $c_1$, $c_2$ are extracted from fits explained above. Notably the location of the critical point can be determined with less than $\%5$ error even with 11 terms. The error is larger for the crossover slope, $c_1$ and $c_2$, however it is below $\%10$  for all cases except for $c_2$ with 11 terms ($\approx \%21$).

\section{Summary and Conclusions}
\label{sec:conclusions}

In this letter we tackled a general problem: given a  truncated series expansion of the equation of state of a theory obtained away from any critical point, what can we say about a the existence of a critical point in the phase diagram and the singular behavior of the equation of state around it?  
We considered a fairly common case of a first order transition that ends at a second order critical point (which is in the Ising universality class) and turns into smooth crossover.  We introduced a new way of extracting the leading singular behavior of the equation of state around the Lee-Yang edge singularity in the complex $\mu$ plane by using a combination of a conformal map and Pad\'e resummation. Equipped with the knowledge of this trajectory, we showed how to extract the location of the critical point and to constrain the non-universal Ising mapping parameters in the scaling region. Finally we demonstrated that this method significantly improves the numerical evaluation of the equation of state even with a modest number of Taylor coefficients. 

There are various extensions and refinements of these ideas that we left for future work. For example, one can extend this analysis to the first order side, $T<T_c$ where one expects a the scaling part of the equation of state to jump to a different Riemann sheet \cite{Fonseca:2001dc, An:2017brc}. This can be achieved by using the so-called ``uniformization map" as a part of conformal Pad\'e which gives access to higher sheets. Another possible application is be to study the Fonseca-Zamolodchikov conjecture this way \cite{Fonseca:2001dc}.  How to handle various uncertainties in the original Taylor coefficients is another non-trivial problem that is currently under investigation. 

An immediate application of the ideas we developed in this letter would be to assist the search for the critical point of QCD both by constraining its location and the Ising mapping parameters, as well as improving the numerical implementation of the equation of state in the hydrodynamic simulations to smoothly interpolate the small $\mu$ behavior to the scaling region. Another possible extension of this work is to incorporate it with analytical continuation from imaginary $\mu$, especially with the recently introduced resummation method \cite{Borsanyi:2021sxv}. 

\acknowledgments
\section{Acknowledgments}

We thank G. Dunne and M. Stephanov for valuable discussions.

\bibliographystyle{utphys}
\bibliography{references}

\providecommand{\href}[2]{#2}\begingroup\raggedright\begin{thebibliography}{10}

\bibitem{Fisher:1974series}
M.~E. Fisher, ``Critical point phenomena - the role of series expansions,''
  {\em Rocky Mountain Journal of Mathematics} {\bfseries 4} no.~2, (1974) 181.

\bibitem{Stephanov:2004wx}
M.~A. Stephanov, ``{QCD phase diagram and the critical point},''
  \href{http://dx.doi.org/10.1142/S0217751X05027965}{{\em Prog. Theor. Phys.
  Suppl.} {\bfseries 153} (2004) 139--156},
  \href{http://arxiv.org/abs/hep-ph/0402115}{{\ttfamily arXiv:hep-ph/0402115}}.

\bibitem{Bzdak:2019pkr}
A.~Bzdak, S.~Esumi, V.~Koch, J.~Liao, M.~Stephanov, and N.~Xu, ``{Mapping the
  Phases of Quantum Chromodynamics with Beam Energy Scan},''
  \href{http://dx.doi.org/10.1016/j.physrep.2020.01.005}{{\em Phys. Rept.}
  {\bfseries 853} (2020) 1--87},
  \href{http://arxiv.org/abs/1906.00936}{{\ttfamily arXiv:1906.00936
  [nucl-th]}}.

\bibitem{Ratti:2018ksb}
C.~Ratti, ``{Lattice QCD and heavy ion collisions: a review of recent
  progress},'' \href{http://dx.doi.org/10.1088/1361-6633/aabb97}{{\em Rept.
  Prog. Phys.} {\bfseries 81} no.~8, (2018) 084301},
  \href{http://arxiv.org/abs/1804.07810}{{\ttfamily arXiv:1804.07810
  [hep-lat]}}.

\bibitem{Liu_2013}
X.-J. Liu, ``Virial expansion for a strongly correlated fermi system and its
  application to ultracold atomic fermi gases,''
  \href{http://dx.doi.org/10.1016/j.physrep.2012.10.004}{{\em Physics Reports}
  {\bfseries 524} no.~2, (Mar, 2013) 37--83}.
  \url{http://dx.doi.org/10.1016/j.physrep.2012.10.004}.

\bibitem{Yang:1952be}
C.-N. Yang and T.~D. Lee, ``{Statistical theory of equations of state and phase
  transitions. 1. Theory of condensation},''
  \href{http://dx.doi.org/10.1103/PhysRev.87.404}{{\em Phys. Rev.} {\bfseries
  87} (1952) 404--409}.

\bibitem{Lee:1952ig}
T.~D. Lee and C.-N. Yang, ``{Statistical theory of equations of state and phase
  transitions. 2. Lattice gas and Ising model},''
  \href{http://dx.doi.org/10.1103/PhysRev.87.410}{{\em Phys. Rev.} {\bfseries
  87} (1952) 410--419}.

\bibitem{Fisher:1978pf}
M.~E. Fisher, ``{Yang-Lee Edge Singularity and phi**3 Field Theory},''
  \href{http://dx.doi.org/10.1103/PhysRevLett.40.1610}{{\em Phys. Rev. Lett.}
  {\bfseries 40} (1978) 1610--1613}.

\bibitem{An:2017brc}
X.~An, D.~Mesterh\'azy, and M.~A. Stephanov, ``{On spinodal points and Lee-Yang
  edge singularities},'' \href{http://dx.doi.org/10.1088/1742-5468/aaac4a}{{\em
  J. Stat. Mech.} {\bfseries 1803} no.~3, (2018) 033207},
  \href{http://arxiv.org/abs/1707.06447}{{\ttfamily arXiv:1707.06447
  [hep-th]}}.

\bibitem{An:2017rfa}
X.~An, D.~Mesterhazy, and M.~A. Stephanov, ``{Critical fluctuations and complex
  spinodal points},'' \href{http://dx.doi.org/10.22323/1.311.0040}{{\em PoS}
  {\bfseries CPOD2017} (2018) 040}.

\bibitem{ZinnJustin:2002ru}
J.~Zinn-Justin, ``{Quantum field theory and critical phenomena},'' {\em Int.
  Ser. Monogr. Phys.} {\bfseries 113} (2002) 1--1054.

\bibitem{Connelly:2020gwa}
A.~Connelly, G.~Johnson, F.~Rennecke, and V.~Skokov, ``{Universal Location of
  the Yang-Lee Edge Singularity in $O(N)$ Theories},''
  \href{http://dx.doi.org/10.1103/PhysRevLett.125.191602}{{\em Phys. Rev.
  Lett.} {\bfseries 125} no.~19, (2020) 191602},
  \href{http://arxiv.org/abs/2006.12541}{{\ttfamily arXiv:2006.12541
  [cond-mat.stat-mech]}}.

\bibitem{An:2016lni}
X.~An, D.~Mesterh\'azy, and M.~A. Stephanov, ``{Functional renormalization
  group approach to the Yang-Lee edge singularity},''
  \href{http://dx.doi.org/10.1007/JHEP07(2016)041}{{\em JHEP} {\bfseries 07}
  (2016) 041}, \href{http://arxiv.org/abs/1605.06039}{{\ttfamily
  arXiv:1605.06039 [hep-th]}}.

\bibitem{Halasz_1997}
M.~Halasz, A.~Jackson, and J.~Verbaarschot, ``Yang-lee zeros of a random matrix
  model for qcd at finite density,''
  \href{http://dx.doi.org/10.1016/s0370-2693(97)00015-4}{{\em Physics Letters
  B} {\bfseries 395} no.~3-4, (Mar, 1997) 293--297}.
  \url{http://dx.doi.org/10.1016/S0370-2693(97)00015-4}.

\bibitem{Ejiri:2005ts}
S.~Ejiri, ``{Lee-Yang zero analysis for the study of QCD phase structure},''
  \href{http://dx.doi.org/10.1103/PhysRevD.73.054502}{{\em Phys. Rev. D}
  {\bfseries 73} (2006) 054502},
  \href{http://arxiv.org/abs/hep-lat/0506023}{{\ttfamily
  arXiv:hep-lat/0506023}}.

\bibitem{Stephanov:2006dn}
M.~A. Stephanov, ``{QCD critical point and complex chemical potential
  singularities},'' \href{http://dx.doi.org/10.1103/PhysRevD.73.094508}{{\em
  Phys. Rev. D} {\bfseries 73} (2006) 094508},
  \href{http://arxiv.org/abs/hep-lat/0603014}{{\ttfamily
  arXiv:hep-lat/0603014}}.

\bibitem{WAKAYAMA2019227}
M.~Wakayama, V.~Bornyakov, D.~Boyda, V.~Goy, H.~Iida, A.~Molochkov,
  A.~Nakamura, and V.~Zakharov, ``Lee-yang zeros in lattice qcd for searching
  phase transition points,''
  \href{http://dx.doi.org/https://doi.org/10.1016/j.physletb.2019.04.040}{{\em
  Physics Letters B} {\bfseries 793} (2019) 227--233}.
  \url{https://www.sciencedirect.com/science/article/pii/S0370269319302734}.

\bibitem{Mukherjee:2019eou}
S.~Mukherjee and V.~Skokov, ``{Universality driven analytic structure of the
  QCD crossover: radius of convergence in the baryon chemical potential},''
  \href{http://dx.doi.org/10.1103/PhysRevD.103.L071501}{{\em Phys. Rev. D}
  {\bfseries 103} no.~7, (2021) L071501},
  \href{http://arxiv.org/abs/1909.04639}{{\ttfamily arXiv:1909.04639
  [hep-ph]}}.

\bibitem{Connelly:2020pno}
A.~Connelly, G.~Johnson, S.~Mukherjee, and V.~Skokov, ``{Universality driven
  analytic structure of QCD crossover: radius of convergence and QCD critical
  point},'' \href{http://dx.doi.org/10.1016/j.nuclphysa.2020.121834}{{\em Nucl.
  Phys. A} {\bfseries 1005} (2021) 121834},
  \href{http://arxiv.org/abs/2004.05095}{{\ttfamily arXiv:2004.05095
  [hep-ph]}}.

\bibitem{Schmidt:2021pey}
C.~Schmidt, J.~Goswami, G.~Nicotra, F.~Ziesch\'e, P.~Dimopoulos, F.~Di~Renzo,
  S.~Singh, and K.~Zambello, ``{Net-baryon number fluctuations},'' in {\em
  {Criticality in QCD and the Hadron Resonance Gas}}.
\newblock 1, 2021.
\newblock \href{http://arxiv.org/abs/2101.02254}{{\ttfamily arXiv:2101.02254
  [hep-lat]}}.

\bibitem{Parotto:2018pwx}
P.~Parotto, M.~Bluhm, D.~Mroczek, M.~Nahrgang, J.~Noronha-Hostler,
  K.~Rajagopal, C.~Ratti, T.~Sch\"afer, and M.~Stephanov, ``{QCD equation of
  state matched to lattice data and exhibiting a critical point singularity},''
  \href{http://dx.doi.org/10.1103/PhysRevC.101.034901}{{\em Phys. Rev. C}
  {\bfseries 101} no.~3, (2020) 034901},
  \href{http://arxiv.org/abs/1805.05249}{{\ttfamily arXiv:1805.05249
  [hep-ph]}}.

\bibitem{Pradeep:2019ccv}
M.~S. Pradeep and M.~Stephanov, ``{Universality of the critical point mapping
  between Ising model and QCD at small quark mass},''
  \href{http://dx.doi.org/10.1103/PhysRevD.100.056003}{{\em Phys. Rev. D}
  {\bfseries 100} no.~5, (2019) 056003},
  \href{http://arxiv.org/abs/1905.13247}{{\ttfamily arXiv:1905.13247
  [hep-ph]}}.

\bibitem{PhysRevD.99.114510}
M.~Giordano and A.~P\'asztor, ``Reliable estimation of the radius of
  convergence in finite density qcd,''
  \href{http://dx.doi.org/10.1103/PhysRevD.99.114510}{{\em Phys. Rev. D}
  {\bfseries 99} (Jun, 2019) 114510}.
  \url{https://link.aps.org/doi/10.1103/PhysRevD.99.114510}.

\bibitem{GIORDANO2021121986}
M.~Giordano, K.~Kapas, S.~Katz, D.~Nogradi, and A.~Pasztor, ``Towards a
  reliable lower bound on the location of the critical endpoint,''
  \href{http://dx.doi.org/https://doi.org/10.1016/j.nuclphysa.2020.121986}{{\em
  Nuclear Physics A} {\bfseries 1005} (2021) 121986}.
  \url{https://www.sciencedirect.com/science/article/pii/S0375947420302967}.
  The 28th International Conference on Ultra-relativistic Nucleus-Nucleus
  Collisions: Quark Matter 2019.

\bibitem{Costin:2020pcj}
O.~Costin and G.~V. Dunne, ``{Uniformization and Constructive Analytic
  Continuation of Taylor Series},''
  \href{http://arxiv.org/abs/2009.01962}{{\ttfamily arXiv:2009.01962
  [math.CV]}}.

\bibitem{Guida:1998bx}
R.~Guida and J.~Zinn-Justin, ``{Critical exponents of the N vector model},''
  \href{http://dx.doi.org/10.1088/0305-4470/31/40/006}{{\em J. Phys. A}
  {\bfseries 31} (1998) 8103--8121},
  \href{http://arxiv.org/abs/cond-mat/9803240}{{\ttfamily
  arXiv:cond-mat/9803240}}.

\bibitem{Serone:2019szm}
M.~Serone, G.~Spada, and G.~Villadoro, ``{$\lambda \phi_2^4$ theory
  \textemdash{} Part II. the broken phase beyond NNNN(NNNN)LO},''
  \href{http://dx.doi.org/10.1007/JHEP05(2019)047}{{\em JHEP} {\bfseries 05}
  (2019) 047}, \href{http://arxiv.org/abs/1901.05023}{{\ttfamily
  arXiv:1901.05023 [hep-th]}}.

\bibitem{Rossi:2018}
R.~Rossi, T.~Ohgoe, K.~Van~Houcke, and F.~Werner, ``Resummation of diagrammatic
  series with zero convergence radius for strongly correlated fermions,''
  \href{http://dx.doi.org/10.1103/PhysRevLett.121.130405}{{\em Phys. Rev.
  Lett.} {\bfseries 121} (Sep, 2018) 130405}.
  \url{https://link.aps.org/doi/10.1103/PhysRevLett.121.130405}.

\bibitem{Gross:1974jv}
D.~J. Gross and A.~Neveu, ``{Dynamical Symmetry Breaking in Asymptotically Free
  Field Theories},'' \href{http://dx.doi.org/10.1103/PhysRevD.10.3235}{{\em
  Phys. Rev. D} {\bfseries 10} (1974) 3235}.

\bibitem{Chodos:1993mf}
A.~Chodos and H.~Minakata, ``{The Gross-Neveu model as an effective theory for
  polyacetylene},'' \href{http://dx.doi.org/10.1016/0375-9601(94)90557-6}{{\em
  Phys. Lett. A} {\bfseries 191} (1994) 39}.

\bibitem{Schnetz:2004vr}
O.~Schnetz, M.~Thies, and K.~Urlichs, ``{Phase diagram of the Gross-Neveu
  model: Exact results and condensed matter precursors},''
  \href{http://dx.doi.org/10.1016/j.aop.2004.06.009}{{\em Annals Phys.}
  {\bfseries 314} (2004) 425--447},
  \href{http://arxiv.org/abs/hep-th/0402014}{{\ttfamily arXiv:hep-th/0402014}}.

\bibitem{Schnetz:2005ih}
O.~Schnetz, M.~Thies, and K.~Urlichs, ``{Full phase diagram of the massive
  Gross-Neveu model},'' \href{http://dx.doi.org/10.1016/j.aop.2005.12.007}{{\em
  Annals Phys.} {\bfseries 321} (2006) 2604--2637},
  \href{http://arxiv.org/abs/hep-th/0511206}{{\ttfamily arXiv:hep-th/0511206}}.

\bibitem{Basar:2009fg}
G.~Basar, G.~V. Dunne, and M.~Thies, ``{Inhomogeneous Condensates in the
  Thermodynamics of the Chiral NJL(2) model},''
  \href{http://dx.doi.org/10.1103/PhysRevD.79.105012}{{\em Phys. Rev. D}
  {\bfseries 79} (2009) 105012},
  \href{http://arxiv.org/abs/0903.1868}{{\ttfamily arXiv:0903.1868 [hep-th]}}.

\bibitem{Fonseca:2001dc}
P.~Fonseca and A.~Zamolodchikov, ``{Ising field theory in a magnetic field:
  Analytic properties of the free energy},''
  \href{http://arxiv.org/abs/hep-th/0112167}{{\ttfamily arXiv:hep-th/0112167}}.

\bibitem{Borsanyi:2021sxv}
S.~Bors\'anyi, Z.~Fodor, J.~N. Guenther, R.~Kara, S.~D. Katz, P.~Parotto,
  A.~P\'asztor, C.~Ratti, and K.~K. Szab\'o, ``{Lattice QCD equation of state
  at finite chemical potential from an alternative expansion scheme},''
  \href{http://arxiv.org/abs/2102.06660}{{\ttfamily arXiv:2102.06660
  [hep-lat]}}.

\end{thebibliography}\endgroup

\end{document}